\documentclass[twocolumn,twoside,slac]{revtex4}
\usepackage{graphicx}
\usepackage{fancyhdr}
\begin{document}

\title{The Event as an Object-Relational Database: Avoiding the Dependency Nightmare}


\author{C. D. Jones}
\affiliation{Cornell University, Ithaca, NY 14853, USA}

\begin{abstract}
With the use of object-oriented languages for HEP, many experiments have designed their data objects to contain direct references to other objects in the event (e.g., tracks and electromagnetic showers have references to each other to denote matches).  Unfortunately this creates tremendous dependencies between packages which lead to brittle development systems (e.g., if the electromagnetic code has a problem you may not be able to compile the tracking code) and makes the storage system more complex.

We discuss how the CLEO III experiment avoided these problems by treating an event as an object-relational database.  The discussion will include: the constraints we placed on our objects; our use of a separate Association class to deal with inter-object references; and our ability to use multiple sources to supply different data items for one event.
\end{abstract}

\maketitle

\thispagestyle{fancy}


\section{INTRODUCTION}

Data within an event often relate to one another, e.g., tracks are often matched to showers in the electro-magnetic calorimeter.  A simple object-oriented design for objects in an event has these data items containing pointers to one another.  Unfortunately, this causes serious dependency problems: large compilation times, extremely long link times, and broken code affects more systems.  Using an object-relational model avoids these problems and allows new possibilities.

\section{SIMPLE OBJECT-ORIENTED APPROACH}
\begin{figure*}[t]
\centering
\includegraphics[width=135mm]{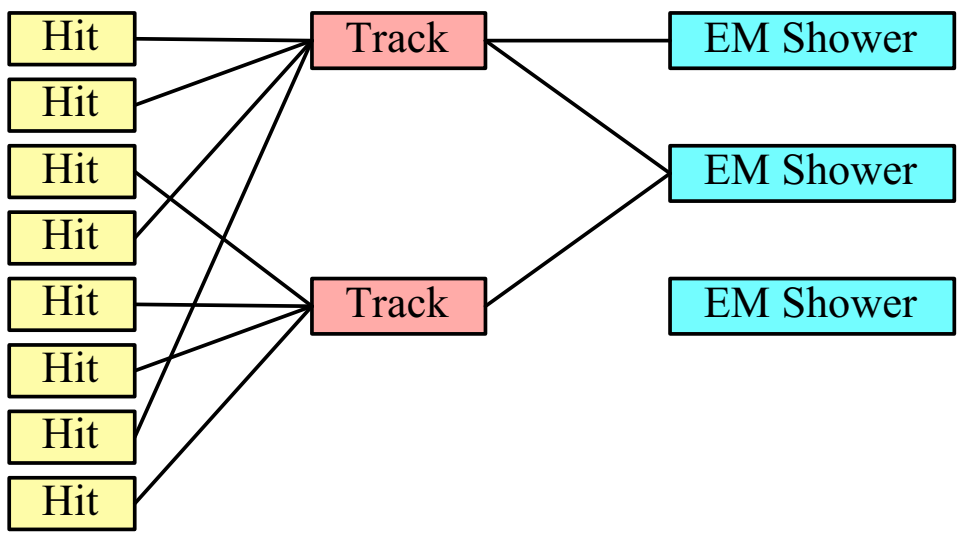}
\caption{Example of simple object-oriented approach to object relationships. Each box represents an object and the lines connecting the boxes represent pointers linking the objects.} \label{oo}
\end{figure*}

A simple object-oriented design for data in an event is shown in Figure~\ref{oo}.  This design has related data grouped into classes, e.g., Track and EM Shower classes.  The data values are stored in objects of the appropriate class. For example, data describing one track is stored in one object of class type Track.  Finally, links between objects are implemented by embedding pointers into the objects.  For example, EM Shower objects hold pointers to any Track which is believed to be matched with that particular shower.  The appeal of this design is that it is easy for users to navigate the relationships between objects.

Unfortunately, the simple object-oriented approach leads to many problems.  The first problem arises in the software's interfaces.  Adding new relationships between objects means changing the classes involved.  This change forces a recompilation of all code that uses those classes.  In addition, published objects (objects that are made available to other parts of the code by placing the objects into the event) must be mutable.  This is because we need to be able to change an object in order to set its relationship to another object.  This design also raises the question of how to handle links in the case where multiple algorithms produce the same data. For example, tracks from different track-finders could be matched to the same EM Showers.  Another question is where to put the data that describes the relationship.  In the "track-shower" matching example, where does the distance between the track and the shower live?  A final question is, how do two people refer to the same object if each has made a sub-selection of a list?  This often arises when people are comparing results from two different analyses that are looking for the same decay mode.

Another set of problems arise for compilation and linking of code.  In highly coupled systems, if one piece of code breaks, the whole system can break.  E.g., if tracking is broken a user may not be able to do EM shower work.  There are standard ways to decrease dependencies in C or C++~\cite{large-scale-cpp} but the techniques may not be known by all developers.  To avoid excess compilation dependencies in C or C++, you must forward declare data in the header files rather than including the data object header files. To avoid excess linking dependencies, associated objects can not internally access member functions of each other.  E.g., we can not have a function that calculates the energy of an EM shower divided by momentum of the track.  It is possible to relax this requirement if you organize your code so that the associated routines are in a separate object file.  This works since many linkers force resolution of all symbols found in an object file.  A further complication is that reference counting smart pointers cause strong compile and link-time dependencies which is unfortunate since they make memory management easier.

The last set of problems we will discuss occur in object storage.  Direct references in objects complicate storage.  This arises since the storage system needs to convert pointers to and from persistent values.  If objects use bidirectional links, it is necessary to construct both objects before linking them.  To simplify the storage system, developers often couple their objects directly to the storage system.  Unfortunately, this coupling locks the developer into using only one storage mechanism even if that mechanism is not appropriate for all the experiment's data.  Reading and writing objects causes compile, link and runtime dependencies between classes.  This is true even if objects only hold pointers to other types of objects.  It is possible to avoid some of these dependencies if the developer is willing to read back unlinked objects.  Unfortunately, use of such unlinked objects forces physicists who use the system to tell the system when the links should be made.  So the user is burdened with the responsibility to be sure the link is made before she tries to use the link.

\section{OBJECT-RELATIONAL APPROACH}

The problems mentioned in the previous section led us to try an object-relational approach.  In this approach, no objects have pointers to objects outside 'atomic' storage boundaries.  E.g., MC particles can hold pointers to their children if they are stored as a unit.  A second requirement of this approach is that all objects in lists must have a unique identifier.  Physicists can use the identifier when talking with other physicists about the objects.  In our system, we use our own templated Table class to hold lists of objects which sort the objects via their identifier method.  Also in our system, lists are identified via unique keys based on the type of the objects in the list plus two character strings.  Therefore objects can be uniquely identified by what list it is in and by what identifier it has within the list. The final requirement of the object-relational approach is to define relationships via separate objects, which we call Lattices~\cite{lattice}.

A Lattice is an object which links relationship data (e.g., the distance between a Track and an EM Shower) to the identifiers of two different objects (denoted by Left and Right).  The Lattice supports all 16 possible configurations for links.  A configuration is defined by four separate sub-configurations where each sub-configuration has two choices.  The four sub-configurations are:
\begin{itemize}
\item 1 or many Lefts per Link
\item 1 or many Rights per Link
\item 1 or many Links per Left
\item 1 or many Links per Right
\end{itemize}

\begin{figure*}[t]
\centering
\includegraphics[width=135mm]{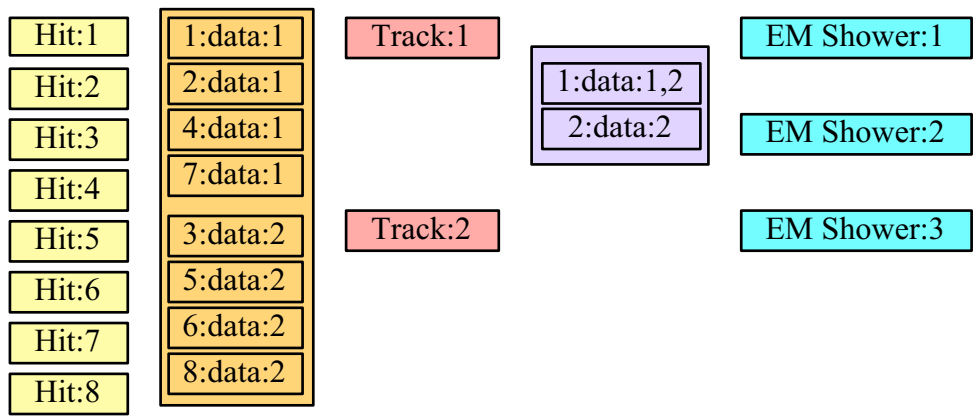}
\caption{Example of the object-relational approach to object relationships.  Each box represents an object.  The boxes within the longer boxes are the Link objects held within each Lattice.  Relationships between objects are determined by matching the numbers held in the Link objects with the appropriate objects identification number.} \label{or}
\end{figure*}

Figure~\ref{or} shows an example of the object-relational approach. The figure shows the Hit, Track and EM Shower objects with each object having a unique identifier within its respective list.  Between the Hit and Track lists is the Hit-to-Track Lattice.  Within the Lattice you can see the separate objects holding the different link information.  For example, the first link in the Hit-to-Track Lattice says that Hit 1 (denoted by the left hand number) is connected with Track 1 (denoted by the right hand number).  Similarly, the Track to EM Shower Lattice is shown between the Track and EM Shower lists.

The object-relational approach has many advantages.  First, it shortens link times.  In our experience, linking usually takes less than 30 seconds on a moderately powerful machine.  However, we use dynamic loading so we only have to link to the libraries a module directly needs and this reduction in the number of libraries needed for linking also contributes to our short link times and allows compilation on very moderate machines.  Second, this approach simplifies the code used for storage.  Because the system is not coupled to one storage mechanism, it is easy to support many specialized storage formats.  Third, this approach speeds up data read-back since we only retrieve data a user actually uses.  E.g., we can ask if a Track is matched to an EM Shower without needing to construct the EM Showers.  Fourth, it is possible to use multiple data sources (each with their own format) on read-back.  E.g., our system can build an event by combining a physicist's data skim with the experiment's event database.

One disadvantage of this approach is it makes navigating the relationships between objects more complex.  To offset this disadvantage, we have created 'Navigation' objects that give direct access to related objects.  These objects internally look up the relationship information in the appropriate Lattice and then use the regular data access mechanism to retrieve the appropriate related objects.  Effectively, the Navigation objects do what users would have to do in order to obtain related objects.   To avoid interdependencies in critical software, only analysis code is allowed to use Navigation objects.  Additionally, we have taken special care so that only by accessing an object via Navigation does the users code become compile/link-time dependent on that object.  E.g., if a user does not use EM Showers then you do not need to link to them, even though the Navigation tracks could access them.  Finally, to make code maintenance easier, we only allow the one library that holds the Navigation objects to have interdependencies between objects.  It also means that only the developer in charge of the Navigation library has to be an expert on how to minimize interdependencies in C++ code.

\section{CONCLUSION}

The general wisdom when writing code is that compile/link/run-time dependencies make code less robust. We have found it possible to avoid unnecessary dependencies by encapsulating relationships between objects into a separate object.  However, providing direct link objects only to analysis users works well. Their code usually accesses most high level data objects, and analysis code has the shortest usage lifetime so long-term maintenance issues are less important.  By following the object-relational approach, we have seen our user's productivity and satisfaction increase because they have gained shorter compile and run times.

\begin{acknowledgments}
This work was supported by the National Science Foundation.
\end{acknowledgments}


\end{document}